\documentclass[epj,referee]{svjour}
\usepackage{amsmath}
\usepackage{graphicx}
\usepackage{color}
\usepackage[english]{babel}
\usepackage{mathpazo}
\newcommand{\thgg}{\theta_{\gamma\gamma\mathrm{cm}}}

\newcommand{\qcm}{q_{\mathrm{cm}}}
\newcommand{\qpcm}{{q'_{\mathrm{cm}}}}

\newcommand{\ammiss}{M_{\mathrm{X}}^2}

\newcommand{\figref}[1]{\ref{#1}}

\newcommand{\degrees}{^{\circ}}
\newcommand{\diff}{\mathrm{d}}
\begin{document}
\title{     
{A new measurement of the structure functions $P_{\mathrm{LL}}-P_{\mathrm{TT}}/\varepsilon$ and $P_{\mathrm{LT}}$ in virtual Compton scattering at $\mathbf Q^2=$ 0.33 (GeV/c)$^2$}
}
\author{
The A1 Collaboration\\[3mm]
P.~Janssens\inst{1} \and L.~Doria\inst{2} \and P.~Achenbach\inst{2}  \and C.~Ayerbe~Gayoso\inst{2}  \and D.~Baumann\inst{2} \and J.C.~Bernauer\inst{2} \and I.K.~Bensafa\inst{3} \and R.~B\"ohm\inst{2} \and D.~Bosnar\inst{4} \and E.~Burtin\inst{5} \and N.~D'Hose\inst{5} \and X.~Defa\"y\inst{3} \and M.~Ding\inst{2} \and M.O.~Distler\inst{2} \and H.~Fonvieille\inst{3}\thanks{e-mail: helene@clermont.in2p3.fr} \and J.~Friedrich\inst{2} \and J.M.~Friedrich\inst{6} \and G.~Laveissi\`ere\inst{3} \and M.~Makek\inst{4} \and J.~Marroncle\inst{5} \and H.~Merkel\inst{2} \and U.~M\"uller\inst{2}  \and L.~Nungesser\inst{2} \and B.~Pasquini\inst{7} \and J.~Pochodzalla\inst{2} \and O.~Postavaru \inst{10} \and M.~Potokar\inst{8} \and D.~Ryckbosch\inst{1} \and S.~Sanchez~Majos\inst{2} \and B.S.~Schlimme\inst{2} \and M.~Seimetz\inst{5} \and S.~\v{S}irca\inst{8,} \inst{9} \and G.~Tamas\inst{2} \and R.~Van~de~Vyver\inst{1} \and L.~Van~Hoorebeke\inst{1} \and A.~Van~Overloop\inst{1} \and Th.~Walcher\inst{2} \and M.~Weinriefer\inst{2} }
\institute{Department of Subatomic and Radiation Physics, University of Gent, 9000 Gent, Belgium. \and Institut f\"ur Kernphysik, Johannes Gutenberg-Universit\"at, 55099 Mainz, Germany. \and LPC, Universit\'e Blaise Pascal, IN2P3, 63177 Aubi\`ere Cedex, France. \and Department of Physics, University of Zagreb, 10002 Zagreb, Croatia. \and CEA DAPNIA-SPhN, C.E. Saclay, 91191 Gif-sur-Yvette Cedex, France. \and Physik-Department, Technische Universit\"at M\"unchen, 85748 Garching, Germany. \and Dipartimento di Fisica Nucleare e Teorica, Universita degli Studi di Pavia, and INFN, Sezione di Pavia, Pavia, Italy. \and Jo\v{z}ef Stefan Institute, Ljubljana, Slovenia. \and Dept. of Physics, University of Ljubljana, Slovenia.  \and Institute of Space Science, RO 76900, Bucharest-Magurele, Romania. }

\date{Received: \ldots / Revised version: \ldots}

\abstract{
The cross section of the $ep \to e' p' \gamma$ reaction has been measured at $Q^2 = 0.33$~(GeV/c)$^2$. The experiment was performed using the electron beam of the MAMI accelerator and the standard detector setup of the A1 Collaboration. The cross section is analyzed using the low-energy theorem for virtual Compton scattering, yielding a new determination of the two structure functions $P_{\mathrm{LL}}-P_{\mathrm{TT}}/\varepsilon$ and $P_{\mathrm{LT}}$ which are linear combinations of the  generalized polarizabilities of the proton. We find somewhat larger values than in the previous investigation at the same $Q^2$. This difference, however, is purely due to our more refined analysis of the data. The results tend to confirm the non-trivial $Q^2$-evolution of the generalized polarizabilities and call for more measurements in the low-$Q^2$ region ($\le$  1 (GeV/c)$^2$).
\PACS{
	{13.60.Fz}{Elastic and Compton scattering}   \and
	{14.20.Dh}{Protons and neutrons}             \and
	{25.30.Rw}{Electroproduction reactions}
           }
 }

\titlerunning{A new measurement of the structure functions $P_{\mathrm{LL}}-P_{\mathrm{TT}}/\varepsilon$ and $P_{\mathrm{LT}}$ ... }
\authorrunning{P. Janssens~\emph{et al.}}

\maketitle

\section{Introduction}
\label{intro}

The internal structure of the proton can be studied using virtual Compton scattering (VCS). In this reaction  ($\gamma^{*} p \rightarrow \gamma p'$) a virtual photon $\gamma^{*}$ scatters off the proton $p$ and a real photon $\gamma$ is produced. The VCS reaction below the pion production threshold allows to measure the generalized polarizabilities of the proton (GPs)~\cite{Arenhovel:1974,Guichon:1995,Vanderhaeghen:1997bx,Guichon:1998xv}.  These GPs are functions of     {the photon four-momentum transfer squared $Q^2$ and they describe} the polarizability locally inside the proton on a distance scale indicated by  $Q^2$~\cite{L'vov:2001fz}. Among the six lowest-order dipole  GPs, two are an extension of the polarizabilities $\alpha_{\mathrm{E}}$ and $\beta_{\mathrm{M}}$ obtained in real Compton scattering (RCS)~\cite{OlmosdeLeon:2001zn},  quantifying the deformation of the charge and magnetization distributions inside the proton caused by a static electric or magnetic field, respectively.

\begin{figure}[ht]
\begin{center}
\resizebox{0.45\textwidth}{!}{%
\includegraphics{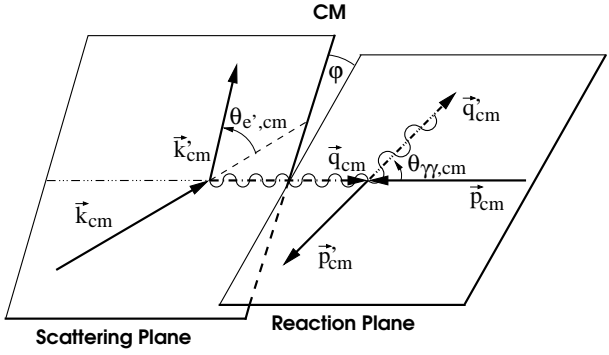}
}
\end{center}
\caption{Schematic drawing of the  $ep \rightarrow e' p' \gamma$ reaction in the center-of-mass of the virtual photon and the target proton.}
\label{fig:vcs}
\end{figure}

VCS  is accessed through photon electroproduction ($ep$ $ \rightarrow$ $ e' p' \gamma$) as shown in figure~\ref{fig:vcs}.  $\vec{k}$, $\vec{p}$ and $\vec{q}$ are the three-momentum vectors of the incoming electron and proton, and the virtual photon, respectively\footnote{All variables defined in the center-of-mass of the virtual photon and target proton have an index `cm'. If no index is given the variable is defined in the laboratory system.}.  $\vec{k}'$, $\vec{p}'$ and $\vec{q}'$ are the three-momentum vectors of the outgoing particles. Five independent variables are necessary to define the kinematics of the reaction, e.g. the modulus of the virtual photon momentum $\qcm$ and its polarization parameter $\varepsilon$, the modulus of the real (outgoing) photon momentum $\qpcm$ and the polar and azimuthal angles of the real photon with respect to the virtual photon direction, $\thgg$ and $\varphi$, respectively. The five-fold differential cross section  $\diff^{5} \sigma / \diff{k' \diff \Omega_{k'} \diff \Omega_{\qpcm}}$ (with $k'=$ modulus of $\vec k '$)  will be noted $\diff^{5} \sigma$.

\begin{figure}[ht]
\begin{center}
\resizebox{0.45\textwidth}{!}{%
\includegraphics{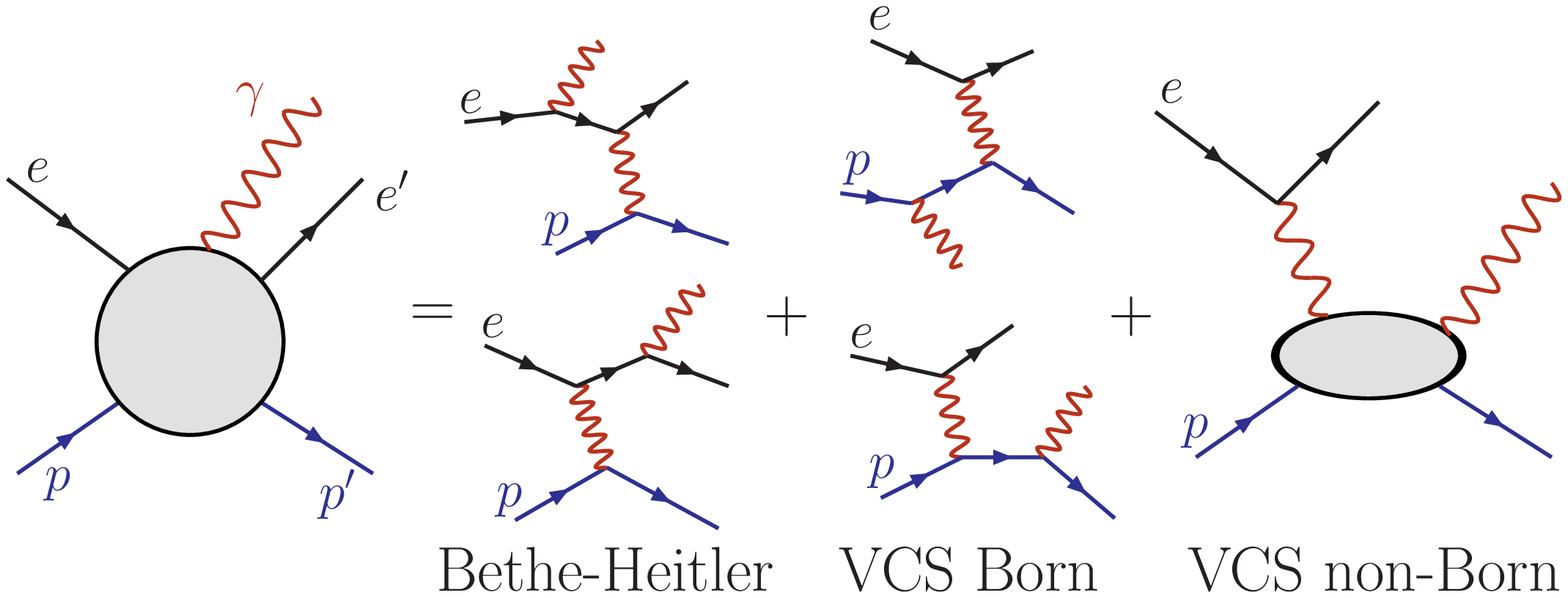}
}
\end{center}
\caption{Contributions to the  $ep \rightarrow e' p' \gamma$ reaction.}
\label{fig:vcs_contr}
\end{figure}

The photon electroproduction reaction contains three contributions (see figure~\ref{fig:vcs_contr}). The reaction is dominated by the Bethe-Heitler and Born (BH+B) contributions, where the outgoing photon is produced due to bremsstrahlung of the electron or proton, respectively. The contribution of the BH+B process can be calculated exactly based on the proton form factors. The GPs make up the VCS non-Born part of the reaction~\cite{Guichon:1995}.

%

\section{Experimental determination of the GPs}
\label{sec:sec2}

The GPs cannot be measured directly. In the physical observables (cross sections and asymmetries) they appear in specific linear combinations, the structure functions. There are six independent GPs, and, by consequence, there are six independent structure functions~\cite{Guichon:1998xv}. Only three of them, denoted as $P_{\mathrm{LL}}$, $P_{\mathrm{TT}}$ and $P_{\mathrm{LT}}$, appear at leading order in the low-energy expansion of the unpolarized cross section. The set of all six GPs can be obtained only by double-polarized VCS. Such an experiment was performed at MAMI recently for the first time~\cite{Merkel:2001} using a longitudinally polarized electron beam and measuring the recoil proton polarization. The analysis of the double polarization asymmetry~\cite{Janssens:2007,Doria:2007}  will be the subject of forthcoming publications. These data can also be used for the determination of the unpolarized cross section by neglecting the beam and recoil proton polarizations. The present paper reports the results of the unpolarized analysis from this experiment~\cite{Janssens:2007}. The same set of structure functions has been determined previously in several experiments at various values of $Q^2$~\cite{Bourgeois:2006js,Roche:2000,Laveissiere:2004}. 

The low-energy theorem (LET) for virtual Compton scattering is used to expand the cross section in powers of $\qpcm$~(\cite{Guichon:1995} and \cite{Guichon:1998xv}):
\begin{equation}
\label{eq:unpol}
\diff^5\sigma = \diff^5 \sigma^{\mathrm{BH+B}} + \phi \qpcm M_0^{\mathrm{NB}} + {\mathcal O}({\qpcm}^2) \; ,
\end{equation}
where $\phi$ is a known phase-space factor, $\diff^5 \sigma^{\mathrm{BH+B}}$ is the five-fold differential cross section for the BH+B processes and $M_0^{\mathrm{NB}}$, which contains the information about the GPs, is defined by
\begin{equation}
\label{eq:linearunpol}
\frac{M_0^{\mathrm{NB}}}{v_{\mathrm{LT}}} = \frac{v_{\mathrm{LL}}}{v_{\mathrm{LT}}} (P_{\mathrm{LL}} - P_{\mathrm{TT}}/\varepsilon) + P_{\mathrm{LT}} \; .
\end{equation}
In this equation $v_{\mathrm{LT}}$ and $v_{\mathrm{LL}}$ are known kinematical functions of $\qcm$, $\varepsilon$, $\thgg$ and $\varphi$ (see e.g. ref.~\cite{Guichon:1998xv} for their complete definition). At fixed $\varepsilon$, two linear combinations of structure functions, $P_{\mathrm{LL}} - P_{\mathrm{TT}}/\varepsilon$ and $P_{\mathrm{LT}}$, can be determined experimentally. The LET method assumes that, since the higher-order terms ${\mathcal O}({\qpcm}^2)$  in eq.~\eqref{eq:unpol} are small for low $\qpcm$ (a condition which holds below the pion production threshold), they can be neglected. The cross section is measured in an appropriate kinematical region, i.e. covering a range large enough in $v_{\mathrm{LL}}$ and $v_{\mathrm{LT}}$, here provided by a large coverage in $\thgg$. Then one forms the quantity $M_0^{\mathrm{NB}} $ = $( \diff^5\sigma - \diff^5 \sigma^{\mathrm{BH+B}} )$ / $( \phi \qpcm )$, which is fitted to a linear combination of the two structure functions, as expressed by eq.~\eqref{eq:linearunpol}.

\section{Experimental setup and event analysis }

For the present experiment the standard setup of the  A1 Collaboration at MAMI was used~\cite{Blomqvist:1998xn} together with the polarized electron beam and the focal plane proton polarimeter. These two items are not detailed here since they play no role in the unpolarized analysis. The beam from the MAMI accelerator impinged with an energy of 854.6 MeV on a liquid hydrogen target. The temperature and pressure inside the target cell were constantly monitored and the beam charge was measured continuously by a F\"orster probe, allowing to determine the experimental luminosity $\mathcal{L}_{\mathrm{exp}}$ with good precision (well below 1\%). The mean beam current was about 22 $\mu$A. To prevent local boiling of the hydrogen, the beam was deflected with an amplitude of a few mm and a frequency of a few kHz. The scattered electron and the recoiling proton were detected in the high resolution spectrometers B and A, respectively. The setting of the spectrometers is given in table~\ref{tab:specsettings}. This setting resulted in the central values  $\qcm=$ 600~MeV/$c$, $\qpcm=$ 90 MeV/$c$, $\varepsilon=$ 0.645 and  $\varphi=180^{\circ}$, and the spectrometer acceptance covered the range [70, 180]$^{\circ}$ for $\thgg$.
  
\begin{table}\centering%
  \caption{Parameters of the spectrometer setup for the VCS90b-setting: central momentum, in-plane angle ($\theta$) and out-of-plane angle ($\phi_{\mathrm{oop}}$). }%
  \begin{tabular}{ c c c c }%
  \hline%
   Parameter & Spectrometer A & Spectrometer B & Unit \\%
  \hline%
  $p_{\mathrm{central}}$ &645.4 & 539.4 & MeV/$c$\\%
  $\theta$  & 38.0 & 50.6 & deg \\%
  $\phi_{\mathrm{oop}}$  & 0.0 & 0.0 & deg \\%
  \hline%
  \end{tabular}%
  \label{tab:specsettings}%
\end{table} %

Each spectrometer contains two sets of double-planes of vertical drift chambers for track reconstruction and two scintillator planes for timing and particle identification. The gas Cherenkov counter in spectrometer B identifies electrons. At the proton side such a Cherenkov detector was not present, since the focal-plane polarimeter was mounted in spectrometer A.  

\begin{figure}[tb]
\begin{center}
\includegraphics[width=0.45\textwidth]{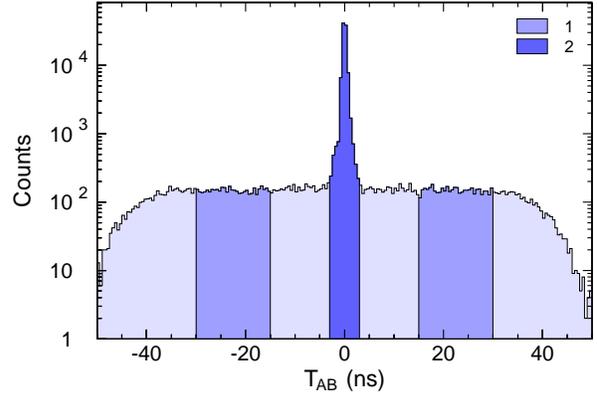}
\end{center}
\caption{Histogram of the coincidence time $T_{\mathrm{AB}}$ for events within the analysis cuts. The coincident events inside the central peak (distribution 2) are selected and distribution 1 is used for the subtraction of random coincidences.}
\label{fig:coincidencetime}
\end{figure}

The distribution of the coincidence time $T_{\mathrm{AB}}$ is shown in figure~\ref{fig:coincidencetime}. The coincident events are selected in a time window of 6~ns and a subtraction of the random coincidences is performed using the events inside $T_{\mathrm{AB}} \in$ [-30, -15] ns and $T_{\mathrm{AB}} \in$ [15, 30] ns. This correction is small since the random coincidences contribute to less than 2\% to the central peak in fig.~\ref{fig:coincidencetime} (note the logarithmic scale). 

\begin{figure}[tb]
\begin{center}
\includegraphics[width=0.45\textwidth]{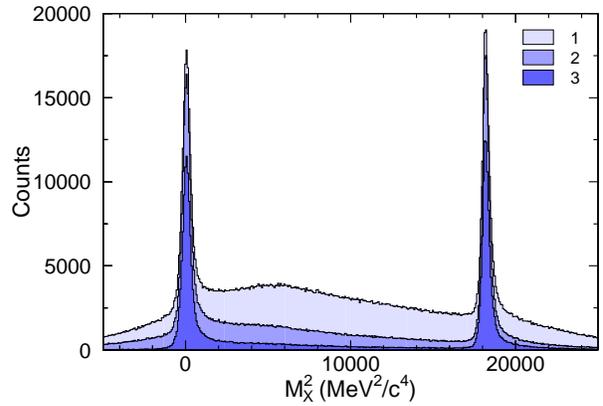}
\end{center}
\caption{Histogram of the square of the missing mass $M_{\mathrm{X}}^2$. Distribution 1 corresponds to the raw coincidences, while in distribution 2 only events within the central peak in $T_{\mathrm{AB}}$  are taken into account and the random coincidences are subtracted. In distribution 3 also the events in the region of the end caps of the target are removed.}
\label{fig:ammissoverview}
\end{figure}

Photon electroproduction events are identified by mis\-sing-mass reconstruction. The distribution of the square of the missing mass $M_{\mathrm{X}}^2$  in  $ep \to ep X$, displayed in figure~\ref{fig:ammissoverview}, shows two peaks corresponding  to $\gamma$ and $\pi^0$ electroproduction. The $\pi^0$  peak is present because the acceptance, while centered on $\qpcm = 90$~MeV/$c$, extends above pion threshold. The separation between the two electroproduction processes 
is excellent (about 30 times the peak FWHM). For the calculation of the cross section only the events with $M_{\mathrm{X}}^2 \in [-1000,2000]$~MeV$^2$/$c^4$ are used. 

A cut on the target length has been applied to remove the events from the interactions of the incoming electrons with the end caps of the target, which are much more dense than the liquid hydrogen itself. Other cuts are necessary to select the events inside the desired kinematic range: $| \qcm - 600$~MeV/$c |$ $< 12$~MeV/$c$, $| \qpcm - 90$~MeV/$c |$ $< 15$~MeV/$c$, $| \varepsilon - 0.645 |$ $< 0.012$ and a range of $\pm 12^{\circ}$ for the out-of-plane angle of the outgoing photon. After these cuts the signals of the scintillators and Cherenkov counters were used to estimate the remaining background processes, which were found to contribute to less than 0.5\%. Since this is well below the statistical uncertainty of the experiment, no cut was applied on these detector signals. The count rates were corrected for the detector efficiency. However, this correction was very small and did not have any influence on the extracted structure functions.

\section{Unpolarized cross section and extraction of structure functions}
\label{sec:unpolcross}

\begin{figure}[hbt]
\begin{center}
\includegraphics[width=0.45\textwidth]{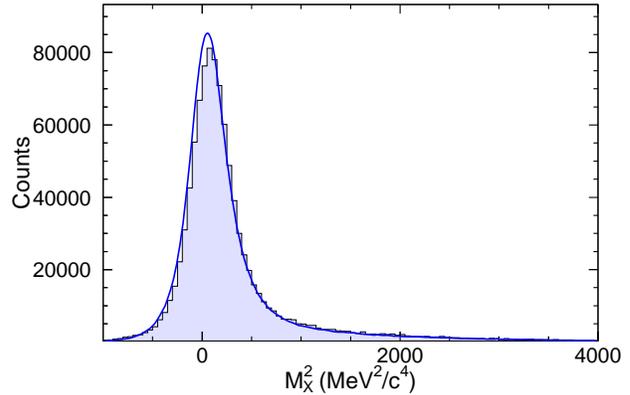}
\end{center}
\caption{Experimental (histogram) and simulated (full line) histogram of $\ammiss$. Both distributions are obtained after applying all cuts. The slight offset in position between both maxima is discussed in the text.}
\label{fig:ammissim}
\end{figure}

For the determination of the cross section the effective solid angle of the detection apparatus is calculated using a Monte Carlo simulation~\cite{Janssens:2006}. This Monte Carlo takes into account the detailed geometry of the apparatus, the beam configuration, and all resolution deteriorating effects, such as the intrinsic resolution of the detectors, energy losses in the materials of the target, etc.
The events are generated according to the BH+B cross section, which is used as an approximation of the real cross section of the photon electroproduction reaction (the non-Born contribution will then be incorporated in an iterative procedure). Radiative effects are taken into account as explained in~\cite{Vanderhaeghen:2000}: the acceptance-dependent part is included in the simulation and the remaining part, due to virtual corrections, is implemented by multiplying the experimental cross section by a constant factor $f_{\mathrm{cor}}$ over the complete phase space. This factor equals 0.942 for the kinematics of the present experiment. The simulation reproduces the radiative tail very well, as can be seen in figure~\ref{fig:ammissim}. 
In this figure the simulated distribution is normalized using the factor $\mathcal{L}_{\mathrm{exp}} / (\mathcal{L}_{\mathrm{sim}} f_{\mathrm{cor}})$, where $\mathcal{L}_{\mathrm{sim}}$ is the luminosity corresponding to the simulated events.  The stability of the experimental cross section versus different $\ammiss$ cuts is better than 1\%. 

The central momenta of the spectrometers were calibrated in absolute using the  $\ammiss$-distribution. By  simultaneously adjusting the experimental peak position (on the simulated one) and minimizing its width, one obtains the two central momenta. This adjustment was done on a kinematical phase space as wide as possible. For the subset of events within the analysis cuts, it results in a slight offset between the experimental and simulated peak positions (fig.~\ref{fig:ammissim}), reflecting the uncertainty of the calibration. This uncertainty is estimated to be $\pm 3 \cdot 10^{-4}$ (in relative) of the central momentum of each spectrometer. It can fully account for the observed offset, i.e. peak positions in fig.~\ref{fig:ammissim} would coincide by changing either one momentum or the other within the quoted uncertainty.

\begin{figure}[htb]
\begin{center}
\includegraphics[width=0.45\textwidth]{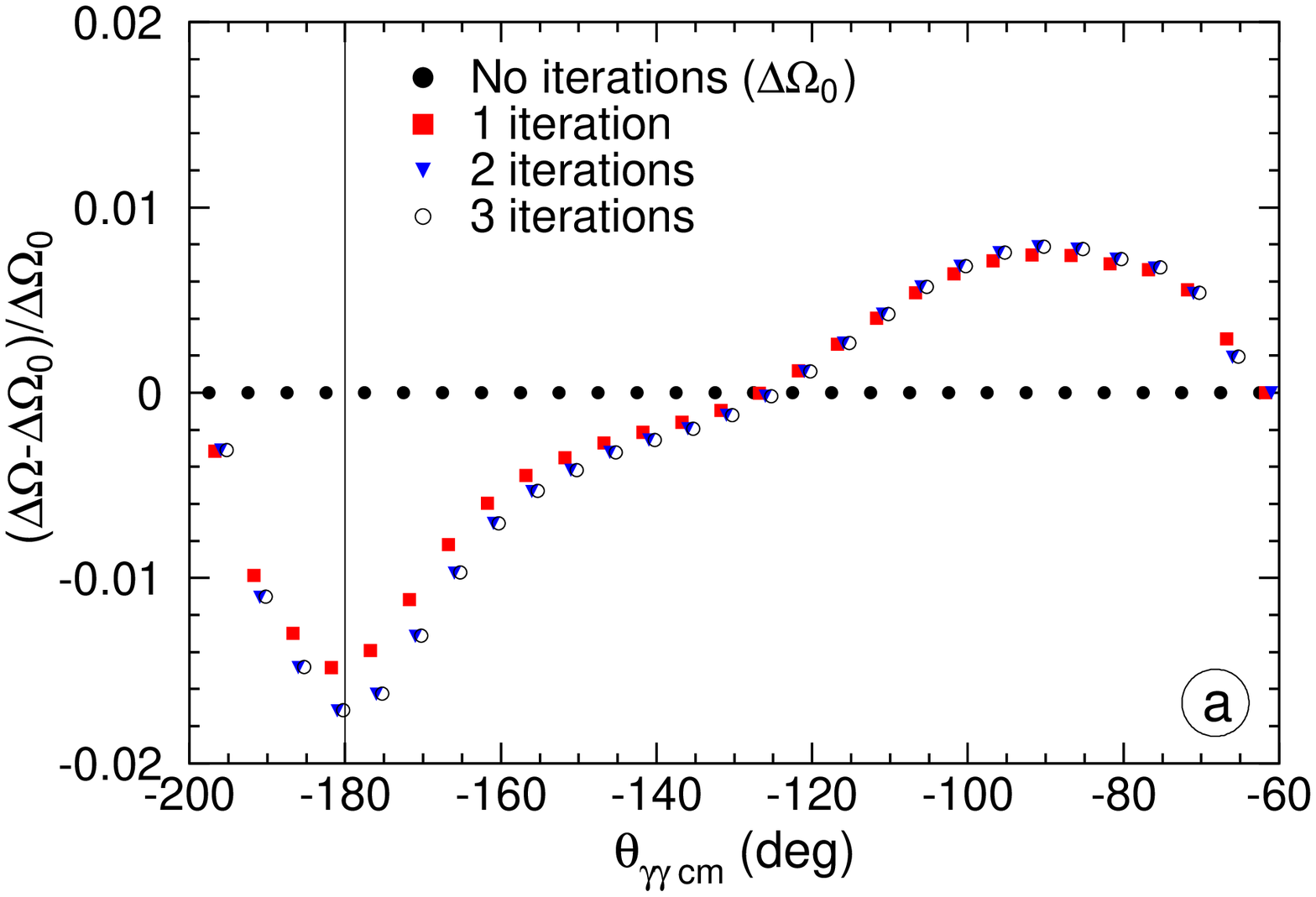}
\includegraphics[width=0.45\textwidth]{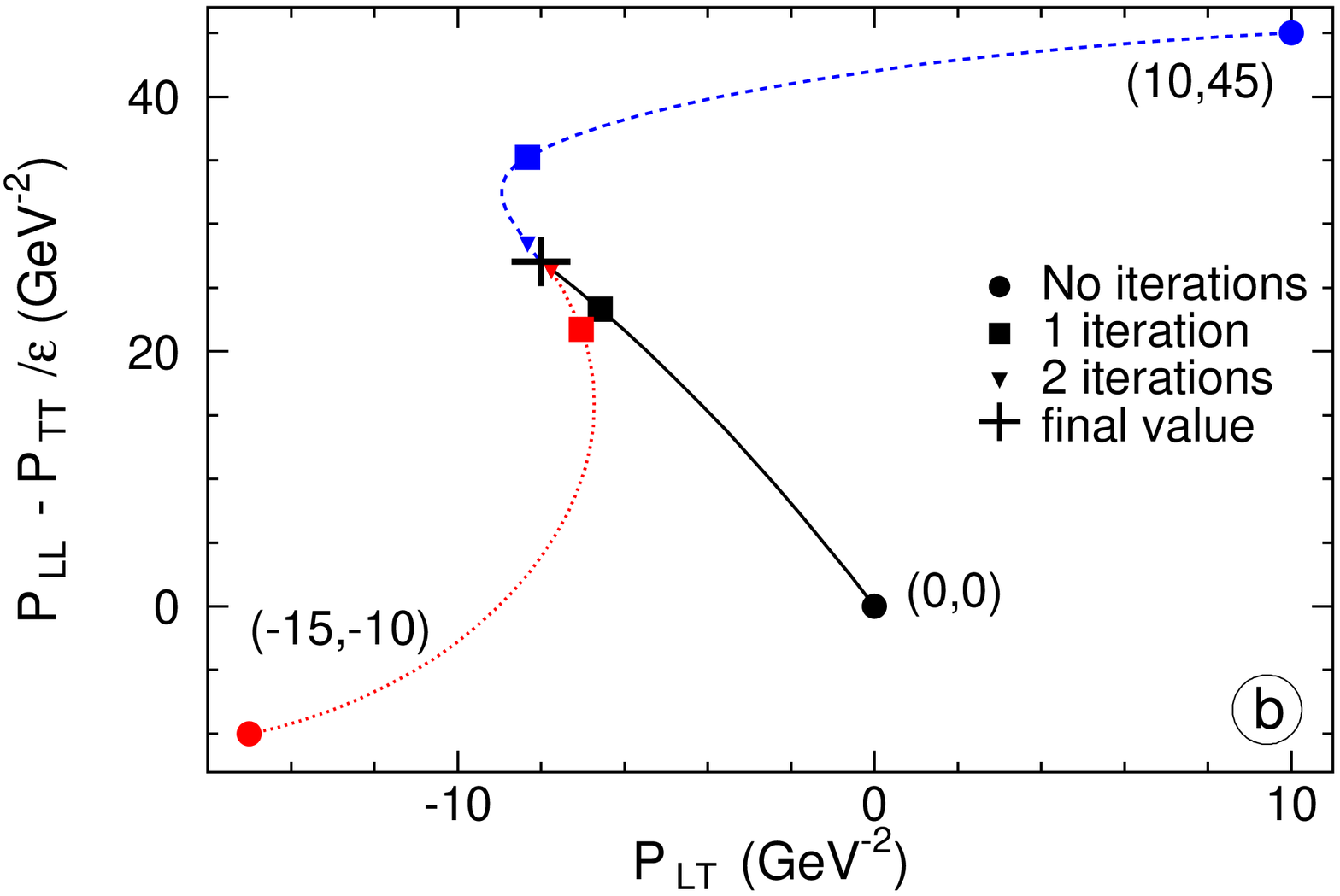}
\end{center}
\caption{
{\bf a}: Influence of the iteration procedure on the effective solid angle. The points at left of  ($\thgg=180^{\circ}$) correspond to $\varphi = 0 \degrees$, the points at right to $\varphi = 180 \degrees$. 
{\bf b}: Convergence pattern for three different starting points, represented by the values of ( $P_{\mathrm{LL}} - P_{\mathrm{TT}}/\varepsilon$, $P_{\mathrm{LT}}$) in parenthesis. The cross at the final point gives the size of the statistical uncertainty. The figure is obtained with the proton form factors of ref.~\cite{Friedrich:2003iz} (see section~\ref{sec-discuss}). 
}
\label{fig:Iterations}
\end{figure}

We now explain the iteration method used to obtain the experimental $(ep \to e' p' \gamma)$ cross section. It is important to use the most realistic cross section for the event generation in the simulation~\cite{Janssens:2006}. In a first step,  using $\diff \sigma^{\mathrm{BH+B}}$ in the simulation, one determines the experimental cross section and extracts the structure functions $P_{\mathrm{LL}} - P_{\mathrm{TT}}/\varepsilon$ and $P_{\mathrm{LT}}$, as explained in section~\ref{sec:sec2}. Then in a second step the cross section in the simulation is modified to include the GP effect measured in the first step. The effective solid angle is recalculated, and the whole procedure is iterated several times. After three iterations a stable result is obtained. The effect of the iterations on the solid angle is shown in figure~\ref{fig:Iterations}.a. The effect is small ($<2\%$), but, due to its pronounced  $\thgg$-dependence, it has a substantial influence on the obtained structure functions (see section~\ref{sec-discuss}). Another feature is that the iteration procedure yields the same result, independent of the initial value for the structure functions. This is represented in figure~\figref{fig:Iterations}.b. When the BH+B cross section is used in the first step of the simulation, the starting point is $(P_{\mathrm{LL}} - P_{\mathrm{TT}}/\varepsilon$, $P_{\mathrm{LT}}) =$ (0,0) GeV$^{-2}$.
 
\begin{table}[h]\centering
  \caption{ The $e p \to e' p' \gamma$ cross section, in pb/(MeV.sr$^2$), obtained in this experiment (after iterations) at fixed  $\qpcm = 90$~MeV/$c$, $\qcm = 600$ MeV/$c$ and $\varepsilon = 0.645$. The error is statistical only. The BH+B cross section is calculated using the form factors of reference~\cite{Friedrich:2003iz}.
}
  \begin{tabular}{ c c c c}
  \hline
 $\thgg$ & $\varphi$  & $\diff^5 \sigma^{\mathrm{BH+B}}$ & $\diff^5 \sigma$\\
\hline
177.5$\degrees$& 180$\degrees$ & 0.129 & 0.146 $\pm$ 0.002\\
172.5$\degrees$& 180$\degrees$ & 0.132 & 0.137 $\pm$ 0.002\\
167.5$\degrees$& 180$\degrees$ & 0.136 & 0.140 $\pm$ 0.002\\
162.5$\degrees$& 180$\degrees$ & 0.142 & 0.148 $\pm$ 0.002\\
157.5$\degrees$& 180$\degrees$ & 0.148 & 0.150 $\pm$ 0.002\\
152.5$\degrees$& 180$\degrees$ & 0.153 & 0.155 $\pm$ 0.002\\
147.5$\degrees$& 180$\degrees$ & 0.156 & 0.154 $\pm$ 0.002\\
142.5$\degrees$& 180$\degrees$ & 0.158 & 0.156 $\pm$ 0.002\\
137.5$\degrees$& 180$\degrees$ & 0.158 & 0.159 $\pm$ 0.002\\
132.5$\degrees$& 180$\degrees$ & 0.157 & 0.157 $\pm$ 0.002\\
127.5$\degrees$& 180$\degrees$ & 0.155 & 0.154 $\pm$ 0.002\\
122.5$\degrees$& 180$\degrees$ & 0.151 & 0.142 $\pm$ 0.002 \\
117.5$\degrees$& 180$\degrees$ & 0.147 & 0.140 $\pm$ 0.002\\
112.5$\degrees$& 180$\degrees$ & 0.142 & 0.135 $\pm$ 0.002\\
107.5$\degrees$& 180$\degrees$ & 0.137 & 0.128 $\pm$ 0.002\\
102.5$\degrees$& 180$\degrees$ & 0.132 & 0.120 $\pm$ 0.003\\
  97.5$\degrees$& 180$\degrees$ & 0.126 & 0.119 $\pm$ 0.003\\
  92.5$\degrees$& 180$\degrees$ & 0.122 & 0.109 $\pm$ 0.003\\
  87.5$\degrees$& 180$\degrees$ & 0.117 & 0.102 $\pm$ 0.003\\
  82.5$\degrees$& 180$\degrees$ & 0.113 & 0.103 $\pm$ 0.003\\
  77.5$\degrees$& 180$\degrees$ & 0.109 & 0.098 $\pm$ 0.003\\
  72.5$\degrees$& 180$\degrees$ & 0.106 & 0.099 $\pm$ 0.005\\
\hline
177.5$\degrees$&     0$\degrees$ & 0.132 & 0.149 $\pm$ 0.003\\
172.5$\degrees$&     0$\degrees$ & 0.141 & 0.167 $\pm$ 0.004\\
\hline
  \end{tabular}
  \label{tab:appcrosssection4}
\end{table}
%

\begin{figure}[htb]
\begin{center}
\includegraphics[width=0.45\textwidth]{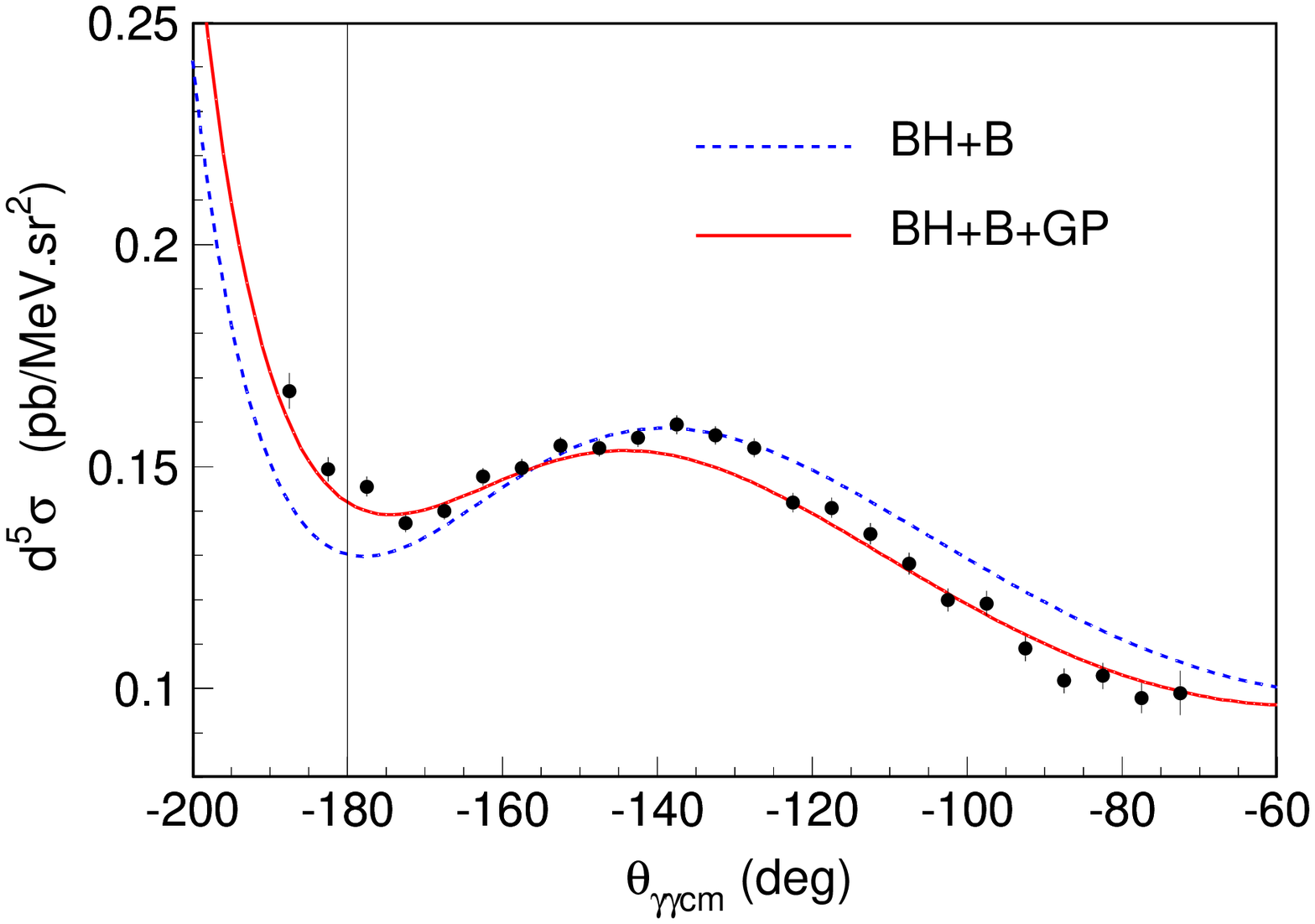}
\end{center}
\caption{The unpolarized $ep \rightarrow e' p' \gamma$ cross section measured at $\qpcm = 90$~MeV/$c$, $\qcm = 600$ MeV/$c$, $\varepsilon = 0.645$ and $\varphi = 180^{\circ}$. The dotted line shows the BH+B cross section, calculated with the proton form factors of ref.~\cite{Friedrich:2003iz}. The full line includes the effect of the structure functions obtained in the experiment (first line of table~\ref{tab:resultiter}). The two most left points correspond to $\varphi = 0 \degrees$.
}
\label{fig:CrossSection}
\end{figure}

The final cross section is displayed in figure~\ref{fig:CrossSection} and the cross section values are given in table~\ref{tab:appcrosssection4}. The main sources of systematic uncertainties are the calibration of the  momenta and angles of the reconstructed particles, the normalization of the cross section (luminosity), the radiative corrections and the simulation of the solid angle. To study the first point, the central momenta of the spectrometers were changed by $\pm 3 \cdot 10^{-4}$ in relative and the data were re-analyzed, yielding new values for the cross section (and the structure functions). The maximal deviation w.r.t. the original values was taken as the uncertainty due to momentum calibration. A procedure along the same lines allows to estimate the systematic error due to the uncertainty in horizontal angle (spectrometer angle plus transfer matrix), taken equal to $\pm 0.1$ mr in each arm. These sources of error are $\thgg$-dependent, changing the shape of the cross section. The three other sources are $\thgg$-independent; summed quadratically, they cause an error in the overall normalization of the cross section of $\pm$ 2\%. The statistical error on the cross section is generally smaller, about $\pm$ 1.4\% for most data points (table~\ref{tab:appcrosssection4}).  

\begin{figure}[htb]
\begin{center}
\includegraphics[width=0.45\textwidth]{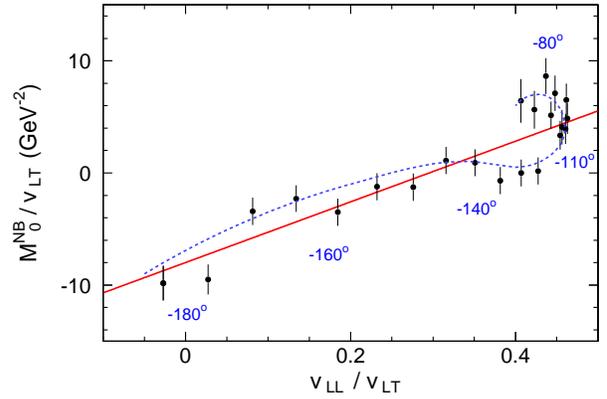}
\end{center}
\caption{ The linear fit of the structure functions: $P_{\mathrm{LL}} - P_{\mathrm{TT}}/\varepsilon$ (slope) and $P_{\mathrm{LT}}$ (intercept). Error bars are statistical only. The angular values reported along the schematic dotted line show the evolution of the points with $\thgg$. The two extreme points ($\thgg$, $\varphi$) = (72.5$^\circ$, 180$^\circ$) and (172.5$^\circ$, 0$^\circ$) are removed from the fit. The figure is obtained with the proton form factors of ref.~\cite{Friedrich:2003iz} and after final iteration.
}
\label{fig:GPExtract}
\end{figure}

The two structure functions  $P_{\mathrm{LL}} - P_{\mathrm{TT}}/\varepsilon$ and $P_{\mathrm{LT}}$ are extracted by a linear fit of the quantity $M_0^{\mathrm{NB}} / v_{\mathrm{LT}}$ as a function of  $v_{\mathrm{LL}}/v_{\mathrm{LT}}$. They are determined at a fixed value of  $\qcm = 600$ MeV/c, or equivalently at a fixed value of $Q^2= 0.33$ (GeV/c)$^2$. The fit is performed via a $\chi^2$ minimization, which also provides the statistical error on the structure functions. The result is shown in figure~\ref{fig:GPExtract}. The reduced $\chi^2$ of the fit, 2.6 for 20 d.o.f.,  suggests that the higher-order terms ${\mathcal O}({\qpcm}^2) $ in eq.~\eqref{eq:unpol}, although small, are not completely negligible. This effect of the LET truncation has not been considered in the error budget. The systematic errors on the structure functions are summarized in table~\ref{tab:SystematicError}, where, as for the cross section, they are separated into an angle-dependent and an angle-independent part (first two lines and  third line, respectively). 

\begin{table}
\caption{Estimation of the systematic error on the structure functions (in GeV$^{-2}$) .}
\label{tab:SystematicError}
\begin{center}
\begin{tabular}{lcc}
\hline\noalign{\smallskip}
 & $P_{\mathrm{LL}} - P_{\mathrm{TT}}/\varepsilon$ & $P_{\mathrm{LT}}$\\
\noalign{\smallskip}\hline\noalign{\smallskip}
Momentum calibration  & $\pm$ 2.7 & $\pm$ 1.0\\
Horizontal angles & $\pm$ 1.2 & $\pm$ 0.4 \\
Normalization of the cross section &  $\pm$ 0.6 &  $\pm$ 1.9\\
\hline
Total systematic error (quadr.sum) &  $\pm$ 3.0 &  $\pm$ 2.2 \\
\noalign{\smallskip}\hline
\end{tabular}
\end{center}
\end{table}

\begin{table}[h]
\caption{Results for $P_{\mathrm{LL}} - P_{\mathrm{TT}}/\varepsilon$ and $P_{\mathrm{LT}}$ without the iterations for $\qcm$~=~600~MeV/$c$ using different form factor parameterizations.
    {The first error is statistical, the second the systematic one.}
$\varepsilon$ in ref.~\cite{Roche:2000} ($\varepsilon = 0.620$) was slightly different from the present experiment ($\varepsilon = 0.645$).}
\label{tab:resultnoiter}
\begin{center}
\begin{tabular}{lccc}
\hline\noalign{\smallskip}
 & $P_{\mathrm{LL}} - P_{\mathrm{TT}}/\varepsilon$& $P_{\mathrm{LT}}$  & Form\\
 &  (GeV$^{-2}$) &  (GeV$^{-2}$) & factor \\
\noalign{\smallskip}\hline\noalign{\smallskip}
This work & 23.3 $\pm$ 1.9 $\pm$ 3.0 &  -6.6 $\pm$ 0.7 $\pm$ 2.2 & \cite{Friedrich:2003iz} \\
& 24.3 $\pm$ 1.9 $\pm$ 3.0 &  -3.9 $\pm$ 0.7 $\pm$ 2.2 & \cite{Mergell:1995bf} \& \cite{Hammer:2003ai}  \\
& 24.7 $\pm$ 1.9 $\pm$ 3.0 & -8.9 $\pm$ 0.7 $\pm$ 2.2  & \cite{Belushkin:2006qa} \\
& 23.7 $\pm$ 1.9 $\pm$ 3.0 &  -5.3 $\pm$ 0.7 $\pm$ 2.2 
& \cite{Hoehler:1976} \& \cite{Kelly:2004hm} \\
\hline
Ref.~\cite{Roche:2000} & 23.7 $\pm$ 2.2 $\pm$ 4.3 &  -5.0 $\pm$ 0.8 $\pm$ 1.8 & \cite{Hoehler:1976}\\
\noalign{\smallskip}\hline
\end{tabular}
\end{center}
\end{table}

\begin{table}[h]
\caption{Results for $P_{\mathrm{LL}} - P_{\mathrm{TT}}/\varepsilon$ and $P_{\mathrm{LT}}$ after applying the iteration procedure for $\qcm$~=~600~MeV/$c$ using different form factor parameterizations.
    {The first error is statistical, the second the systematic one.}
}
\label{tab:resultiter}
\begin{center}
\begin{tabular}{lccc}
\hline\noalign{\smallskip}
 & $P_{\mathrm{LL}} - P_{\mathrm{TT}}/\varepsilon$& $P_{\mathrm{LT}}$  & Form\\
 &  (GeV$^{-2}$) &  (GeV$^{-2}$) & factor\\
\noalign{\smallskip}\hline\noalign{\smallskip}
This work& 27.1 $\pm$ 1.9 $\pm$ 3.0 &  -8.0 $\pm$ 0.7 $\pm$ 2.2 & \cite{Friedrich:2003iz} \\
& 28.5 $\pm$ 1.9 $\pm$ 3.0 &  -5.2 $\pm$ 0.7 $\pm$ 2.2  &  \cite{Mergell:1995bf} \& \cite{Hammer:2003ai}  \\
& 28.6 $\pm$ 1.9 $\pm$ 3.0  &  -10.1  $\pm$ 0.7 $\pm$ 2.2  & \cite{Belushkin:2006qa} \\
& 27.4 $\pm$ 1.9 $\pm$ 3.0  &  -6.8 $\pm$ 0.7 $\pm$ 2.2 & \cite{Hoehler:1976} \& \cite{Kelly:2004hm} \\
\noalign{\smallskip}\hline
\end{tabular}
\end{center}
\end{table}

\section{Discussion} \label{sec-discuss}

The effect of the GPs in the photon electroproduction cross section is small; in the kinematics of the experiment it reaches at maximum $\sim \pm$10\% (see fig.~\ref{fig:CrossSection}). Therefore any small change at the cross section level may induce a relatively large change in the fitted structure functions. 

A first example is provided by the iterative calculation of the solid angle, explained in section~\ref{sec:unpolcross}. Using this procedure, $P_{\mathrm{LL}} - P_{\mathrm{TT}}/\varepsilon$ is increased by 16\% and  $ P_{\mathrm{LT}}$ by 20-30\% (see tables~\ref{tab:resultnoiter} and \ref{tab:resultiter}). In the first VCS experiment at MAMI~\cite{Roche:2000}, this iterative procedure was not pushed to its convergence point, because its effect was smaller than the statistical uncertainty, at the cross section level. Indeed, as can be seen in figure~\ref{fig:Iterations}-a, the iterations change the solid angle (and hence the cross section) by less than 1\% in the main part of the phase space, whereas the statistical uncertainty on the cross section was $\sim$ 2-3\%. Therefore the result of~\cite{Roche:2000} is non-iterated. It can be compared to the non-iterated result of the present analysis, performed at the same value of $\qcm$. As shown in table~\ref{tab:resultnoiter}, at this level the agreement is strikingly good between the two experiments. 

\begin{figure}[htb]
\begin{center}
\includegraphics[width=0.45\textwidth]{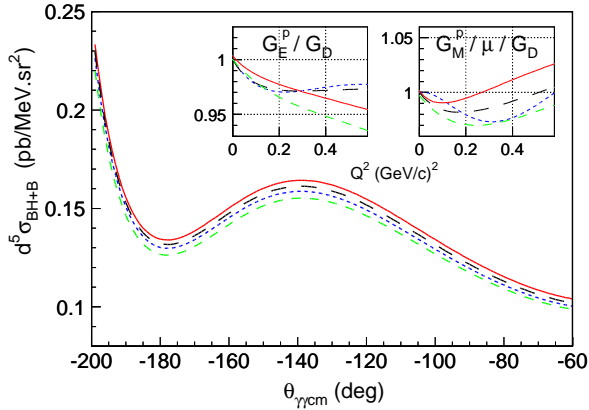} 
\end{center}
\caption{ The BH+B cross section, at the kinematics of the experiment, for proton form factors as parameterized in ~\cite{Friedrich:2003iz} (dotted), 
\cite{Mergell:1995bf}-\cite{Hammer:2003ai} (solid), 
\cite{Belushkin:2006qa} (short-dashed) and \cite{Hoehler:1976} (long-dashed).
The corresponding form factor parameterizations are shown in insert, with $G_{\mathrm{D}}$ = $[ 1+(Q^2/0.71 \mbox{ (GeV/c)}^2 ) ]^{-2} $. Ref.~\cite{Kelly:2004hm} gives curves very similar to~\cite{Hoehler:1976}.
}
\label{fig:protonff}
\end{figure}

A second example is provided by the form factor parameterizations (for recent reviews on nucleon form factors, see refs.~\cite{Arrington:2007ux} and~\cite{HydeWright:2004gh}). The obtained structure functions depend on the choice made for the proton form factors $G_{\mathrm{E}}^{\mathrm{p}}$ and $G_{\mathrm{M}}^{\mathrm{p}}$, because these observables enter the calculation of the  BH+B cross section. In this analysis various form factor parameterizations have been considered \cite{Friedrich:2003iz,Mergell:1995bf,Hammer:2003ai,Belushkin:2006qa,Hoehler:1976,Kelly:2004hm}  (see figure~\ref{fig:protonff}). Mainly $P_{\mathrm{LT}}$ is sensitive to this choice: going from  the parameterization of ref.~\cite{Hoehler:1976} (or \cite{Kelly:2004hm}) to the other ones,  $P_{\mathrm{LT}}$ changes by $ ( ^{+23}_{-48} ) $\%  while $P_{\mathrm{LL}} - P_{\mathrm{TT}}/\varepsilon$ changes by 1-4\% only (cf. table~\ref{tab:resultiter}). This is caused by the fact that a change of form factor induces mainly a variation of the magnitude of the BH+B cross section without affecting too much its $\thgg$-dependence (see figure~\ref{fig:protonff}), resulting mostly in a change of the intercept in figure~\ref{fig:GPExtract}.

It should be noted that in~\cite{Roche:2000} the experimental cross section was compared to the theoretical one also at very low $\qpcm$. This test favored the form factor parameterization of ref.~\cite{Hoehler:1976} which was consequently chosen in the analysis, and an overall  form factor uncertainty was embedded in the systematic error~\footnote{This uncertainty is accounted for in the last line of table~\ref{tab:resultnoiter}.}. The present experiment is only performed at $\qpcm = 90$~MeV/$c$,  preventing such normalization test at low $\qpcm$. The uncertainty due to the proton form factors is not included in the numerical value of the systematic error of table~\ref{tab:resultiter}. It is represented explicitely by the four different lines of results in this table
\footnote{
If however one wants a single number for this systematic error, one may take the half-difference between the two extreme results of table~\ref{tab:resultiter}, i.e. $\pm 0.7$ (resp.$\pm 2.4$) GeV$^{-2}$ for the first (resp.second) structure function. This would give after quadratic sum a total systematic error of  $\pm 3.1$ (resp.$\pm 3.3$) GeV$^{-2}$ on $P_{\mathrm{LL}} - P_{\mathrm{TT}}/\varepsilon$ \ (resp. $P_{\mathrm{LT}}$).
         }.

\begin{figure}[bt]
\begin{center}
\includegraphics[width=0.45\textwidth]{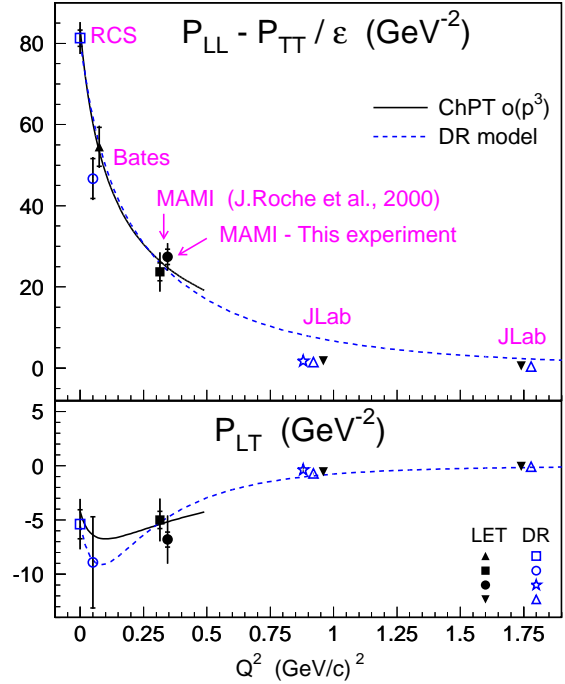} 
\end{center}
\caption{ The structure functions measured in this experiment 
($\bullet$) (with the form factors of ref.~\cite{Hoehler:1976}) and previously at Bates~\cite{Bourgeois:2006js}, MAMI~\cite{Roche:2000}, JLab~\cite{Laveissiere:2004} and in RCS~\cite{OlmosdeLeon:2001zn}. The solid curve is the HBChPT calculation~\cite{Hemmert:1999pz}. The dashed curve is an example of the dispersive (DR) calculation~\cite{Pasquini:2001yy} with dipole ansatz parameters $(\Lambda_{\alpha}, \Lambda_{\beta})=$ (1.8, 0.5) GeV. Curves are calculated for $\epsilon$ = 0.645. Data points correspond to different $\varepsilon$ (0.90~\cite{Bourgeois:2006js}, 0.62~\cite{Roche:2000}, 0.645 (this exp.), 0.95 and 0.88~\cite{Laveissiere:2004}). They are obtained from either a LET analysis or a dispersive one (DR). Some points are shifted in abscissa for visibility. The inner error bar is statistical, the outer one is the quadratic sum of statistical and systematic errors.
}
\label{fig:worldsf}
\end{figure}

The resulting structure functions are displayed in figure~\ref{fig:worldsf} together with the other existing measurements. At low $Q^2$ they can be compared to a calculation in the framework of the heavy baryon chiral perturbation theory (HBChPT)~\cite{Hemmert:1999pz}. The calculation at order $p^3$ for the structure functions evaluated at $Q^2 = 0.33$ (GeV/c)$^2$ and $\varepsilon$ = 0.645 gives  $P_{\mathrm{LL}} - P_{\mathrm{TT}}/\varepsilon = 25.5$~GeV$^{-2}$ and $P_{\mathrm{LT}} = -5.3$~GeV$^{-2}$, in overall good agreement with the values measured in this experiment. At next order in HBChPT, only the spin GPs have been evaluated~\cite{Kao:2002cn,Kao:2004us}. A complete ChPT calculation of the present structure functions, combining scalar and spin GPs, is still to come. The dispersive formalism of refs.~\cite{Pasquini:2001yy,Drechsel:2002ar} offers an alternative approach in which the electric and magnetic GPs can be fitted from the experiment. For a more extensive comparison to the theoretical models we refer the reader to e.g. ref.~\cite{d'Hose:2006xz}. 

In the structure functions presented here, the contribution of the spin GPs (and hence $P_{\mathrm{TT}}$, which contains only spin GPs) is expected to be small. Therefore the data for $P_{\mathrm{LL}} - P_{\mathrm{TT}}/\varepsilon$  essentially  reflect the behavior of $P_{\mathrm{LL}}$, which is proportional to the product $G_{\mathrm{E}}^{\mathrm{p}} \cdot \alpha_{\mathrm{E}} $. The data do not follow a simple dipole shape over the full measured $Q^2$-range. To fix the shape, more measurements are needed in the region up to 1 (GeV/c)$^2$. For the structure function $P_{\mathrm{LT}}$, the scalar part is proportional to $G_{\mathrm{E}}^{\mathrm{p}} \cdot \beta_{\mathrm{M}} $. The measured points tend to confirm the existence of an extremum of  $\beta_{\mathrm{M}}$ at low $Q^2$, traditionally explained by the interplay between diamagnetic and paramagnetic contributions. To summarize, both structure functions show a non-trivial $Q^2$-behavior, that can be related to the pion cloud structure of the nucleon. However, experimental data are still scarce and more measurements in the $Q^2$-region between 0 and 1 (GeV/c)$^2$ would help to gain insight into the matter.

\section{Conclusion}

New values for the structure functions $P_{\mathrm{LL}} - P_{\mathrm{TT}}/\varepsilon$ and $P_{\mathrm{LT}}$ were obtained from the present VCS experiment performed at MAMI. Apart from a well-understood  systematic difference, the results are in very good agreement with the ones obtained in the previous MAMI experiment~\cite{Roche:2000}. They are also in good agreement with the calculation of HBChPT~\cite{Hemmert:1999pz}. The new feature in the present analysis is an iterative procedure in the calculation of the solid angle, inducing an increase of both structure functions. The effect of different proton form factor parameterizations has also been studied in detail. More precise measurements of these form factors at low $Q^2$~\cite{suiteledex,bernauer} will help to reduce the uncertainties in the determination of the GPs. More VCS measurements at low $Q^2$ would help to investigate the non-trivial behavior of the GPs and their connection to the mesonic structure of the nucleon.

\begin{acknowledgement}
We would like to thank the accelerator group of MAMI for its excellent support.
This work was supported in part by the FWO-Flan\-ders (Belgium), the BOF-Gent University, the Deutsche For\-schungsgemeinschaft with the Collaborative Research Center 443, the Federal State of Rhine\-land-Palatinate and the French CEA and CNRS/IN2P3.
\end{acknowledgement}

\bibliography{thesis}

\begin{thebibliography}{10}
\expandafter\ifx\csname url\endcsname\relax
  \def\url#1{\texttt{#1}}\fi
\expandafter\ifx\csname urlprefix\endcsname\relax\def\urlprefix{URL }\fi

\bibitem{Arenhovel:1974}
H.~Arenh{\"o}vel, D.~Drechsel, {Generalized nuclear polarizabilities in
  (e,e'{$\gamma$}) coincidence experiments}, Nucl. Phys. A233 (1974) 153.

\bibitem{Guichon:1995}
P.~A.~M. Guichon, G.~Q. Liu, A.~W. Thomas, {Virtual Compton scattering and
  generalized polarizabilities of the proton}, Nucl. Phys. A591 (1995)
  606--638.

\bibitem{Vanderhaeghen:1997bx}
M.~Vanderhaeghen, {Double polarization observables in virtual Compton
  scattering}, Phys. Lett. B402 (1997) 243--250.

\bibitem{Guichon:1998xv}
P.~A.~M. Guichon, M.~Vanderhaeghen, {Virtual Compton scattering off the
  nucleon}, Prog. Part. Nucl. Phys. 41 (1998) 125--190.

\bibitem{L'vov:2001fz}
A.~I. L'vov, S.~Scherer, B.~Pasquini, C.~Unkmeir, D.~Drechsel, {Generalized
  dipole polarizabilities and the spatial structure of hadrons}, Phys. Rev. C64
  (2001) 015203.

\bibitem{OlmosdeLeon:2001zn}
V.~Olmos~de Leon, et~al., {Low-energy Compton scattering and the
  polarizabilities of the proton}, Eur. Phys. J. A10 (2001) 207--215.

\bibitem{Merkel:2001}
N.~d'Hose, H.~Merkel, {Double Polarization Virtual Compton Scattering in the
  threshold regime at MAMI}, MAMI proposal (2001).

\bibitem{Janssens:2007}
P.~Janssens, Double-polarized virtual compton scattering as a probe of the
  proton structure, Ph.D. thesis, Universiteit Gent (2007).

\bibitem{Doria:2007}
L.~Doria, Polarization observables in virtual compton scattering, Ph.D. thesis,
  Universit{\"a}t Mainz (2008).

\bibitem{Bourgeois:2006js}
P.~Bourgeois, et~al., {Measurements of the generalized electric and magnetic
  polarizabilities of the proton at low $Q^2$ using the VCS reaction}, Phys.
  Rev. Lett. 97 (2006) 212001.

\bibitem{Roche:2000}
J.~Roche, et~al., {The first determination of generalized polarizabilities of
  the proton by a virtual Compton scattering experiment}, Phys. Rev. Lett. 85
  (2000) 708--711.

\bibitem{Laveissiere:2004}
G.~Laveissi{\`e}re, et~al., {Measurement of the generalized polarizabilities of
  the proton in virtual Compton scattering at $Q^2$ = 0.92 GeV$^2$ and 1.76
  GeV$^2$}, Phys. Rev. Lett. 93 (2004) 122001.

\bibitem{Blomqvist:1998xn}
K.~I. Blomqvist, et~al., {The three-spectrometer facility at the Mainz
  microtron MAMI}, Nucl. Instrum. Meth. A403 (1998) 263--301.

\bibitem{Janssens:2006}
P.~Janssens, et~al., {Monte Carlo simulation of virtual Compton scattering
  below pion threshold}, Nucl. Instr. Meth. A566 (2006) 675--686.

\bibitem{Vanderhaeghen:2000}
M.~Vanderhaeghen, et~al., {QED radiative corrections to virtual Compton
  scattering}, Phys. Rev. C62 (2000) 025501.

\bibitem{Friedrich:2003iz}
J.~Friedrich, T.~Walcher, {A coherent interpretation of the form factors of the
  nucleon in terms of a pion cloud and constituent quarks}, Eur. Phys. J. A17
  (2003) 607--623.

\bibitem{Mergell:1995bf}
P.~Mergell, U.~G. Meissner, D.~Drechsel, {Dispersion theoretical analysis of
  the nucleon electromagnetic form-factors}, Nucl. Phys. A596 (1996) 367--396.

\bibitem{Hammer:2003ai}
H.~W. Hammer, U.-G. Meissner, {Updated dispersion-theoretical analysis of the
  nucleon electromagnetic form factors}, Eur. Phys. J. A20 (2004) 469--473.

\bibitem{Belushkin:2006qa}
M.~A. Belushkin, H.~W. Hammer, U.~G. Meissner, {Dispersion analysis of the
  nucleon form factors including meson continua}, Phys. Rev. C75 (2007) 035202.

\bibitem{Hoehler:1976}
G.~H{\"o}hler, et~al., {Analysis of Electromagnetic Nucleon Form Factors},
  Nucl. Phys. B114 (1976) 505--534.

\bibitem{Kelly:2004hm}
J.~J. Kelly, {Simple parametrization of nucleon form factors}, Phys. Rev. C70
  (2004) 068202.

\bibitem{Arrington:2007ux}
J.~Arrington, W.~Melnitchouk, J.~A. Tjon, {Global analysis of proton elastic
  form factor data with two-photon exchange corrections}, Phys. Rev. C76 (2007)
  035205.

\bibitem{HydeWright:2004gh}
C.~E. Hyde-Wright, K.~de~Jager, {Electromagnetic Form Factors of the Nucleon
  and Compton Scattering}, Ann. Rev. Nucl. Part. Sci. 54 (2004) 217--267.

\bibitem{Hemmert:1999pz}
T.~R. Hemmert, B.~R. Holstein, G.~Knochlein, D.~Drechsel, {Generalized
  polarizabilities of the nucleon in chiral effective theories}, Phys. Rev. D62
  (2000) 014013.

\bibitem{Pasquini:2001yy}
B.~Pasquini, M.~Gorchtein, D.~Drechsel, A.~Metz, M.~Vanderhaeghen, {Dispersion
  relation formalism for virtual Compton scattering off the proton}, Eur. Phys.
  J. A11 (2001) 185--208.

\bibitem{Kao:2002cn}
C.~W. Kao, M.~Vanderhaeghen, {Generalized spin polarizabilities of the nucleon
  in heavy baryon chiral perturbation theory at next-to-leading order}, Phys.
  Rev. Lett. 89 (2002) 272002.

\bibitem{Kao:2004us}
C.-W. Kao, B.~Pasquini, M.~Vanderhaeghen, {New predictions for generalized spin
  polarizabilities from heavy baryon chiral perturbation theory}, Phys. Rev.
  D70 (2004) 114004.

\bibitem{Drechsel:2002ar}
D.~Drechsel, B.~Pasquini, M.~Vanderhaeghen, {Dispersion relations in real and
  virtual Compton scattering}, Phys. Rept. 378 (2003) 99--205.

\bibitem{d'Hose:2006xz}
N.~d'Hose, {Virtual Compton scattering at MAMI}, Eur. Phys. J. A28 (2006)
  117--127.

\bibitem{suiteledex}
R.~Gilman, et~al., {Proposal}, JLab PR-07-004 (2006).

\bibitem{bernauer}
M.O.Distler, {Proposal}, MAMI A1-2/05 (2005).

\end{thebibliography}

\end{document}